# IoT based Smart Water Quality Prediction for Biofloc Aquaculture

Md. Mamunur Rashid[1]
Department of CSE, University of Liberal Arts Bangladesh
Dhaka, Bangladesh

Al-Akhir Nayan[2], Md. Obaidur Rahman[5]
Department of CSE, European University of Bangladesh
Dhaka, Bangladesh

Joyeta Saha[4]
Department of ECE, North South University
Dhaka, Bangladesh

Sabrina Afrin Simi[3]
Dept. of Human Computer Interaction
University of Siegen, Siegen, Germany

Muhammad Golam Kibria[6]
Department of CSE, IoT Lab, University of Liberal Arts
Bangladesh, Dhaka, Bangladesh

*Abstract*—**Traditional fish farming faces several challenges, including water pollution, temperature imbalance, feed, space, cost, etc. Biofloc technology in aquaculture transforms the manual into an advanced system that allows the reuse of unused feed by converting them into microbial protein. The objective of the research is to propose an IoT-based solution to aquaculture that increases efficiency and productivity. The article presented a system that collects data using sensors, analyzes them using a machine learning model, generates decisions with the help of Artificial Intelligence (AI), and sends notifications to the user. The proposed system has been implemented and tested to validate and achieve a satisfactory result.**

## I. INTRODUCTION

In biofloc aquaculture, it is inevitable to be more intelligent to monitor the water quality in real-time and feed accurately. However, due to real-time water quality monitoring, the balance of bacteria in the aquaculture environment might be harmed; hence fish's disease-resistant ability is decreased. It is impossible to measure the water quality accurately based on experience only [1, 2]. An intelligent system could help the farmers by reading the water parameters on time to monitor and maintain the quality accordingly. Hence, identifying the water parameters suitable for the biofloc aquaculture, a water quality prediction model for the dynamic changes in water parameters, and accordingly are essential.

Biofloc technology can reduce food costs, while a smart aquaculture system can reduce labor cost. It is a good option that is cheaper and beneficial to fish's health [3, 4]. Being a low-lying country, several natural calamities like floods, cyclones, etc., have a significant effect on aquaculture at both the ponds and marine waters. Fish farmers must bear a substantial loss due to the polluted water and increasing salinity of the coastal water for those disasters. Traditional fish farming leads to several other problems, such as water pollution caused by carbon dioxide, ammonia, and nitrogen. External filtration is needed for detoxifying, which is costly and time-consuming. Biofloc technology is an excellent alternative to the cost-effective traditional aquaculture system. Biofloc itself helps to purify the water naturally, hence the use of external tools or ingredients might be reduced. Maintaining water quality can ensure increasing production, decreasing the death of fish. Water quality parameters are the most important factors to maintain a fish farm using Bioflocs.

This article mainly focuses on water level parameters. An automated system has been implemented to collect data through sensors, analyze them using a machine learning method, analyze the water quality. The application of IoT makes it easier to monitor the water and maintain the ecology in biofloc aquaculture.

In this article, data collected from the fish farm has been used for training and testing purposes. A machine learning method has been applied to develop the model. The water quality parameters, including pH, have been analyzed, and the correlation between them is obtained. The water quality prediction model is trained based on the collected data.

## II. RELATED WORKS

Deep learning (DL) technology is used in numerous fields. X. Yang et al. focused on DL applications in aquaculture. They worked on identifying live and dead fish, classified species, performed behavioral analysis, and feeding decisions. The algorithm and the results of the method were applied to the smart fish farm. The findings showed that the deep learning method could extract features automatically. They made the most valuable contribution to the field of agriculture. But the technique was failed to address complex data in aquaculture [5].

S. Liu et al. did an experiment on "Ras Carpio" using the Recirculating Aquaculture System (RAS) [6]. In 2011, RAS was an intelligent alternative to traditional aquaculture in ponds. The water parameters were being continually monitored, and whenever the parameter's value got out of the fish's versatile range, the water was recirculated. For this purpose, there were two drainage systems. DO, pH, and the temperature was monitored by WATT TriO Matic 700IQ (SW), WATT Sensolyt 700 IQ (SW), and WATT Tri oxyTherm type sensors. Though the system had many benefits and was replaced rapidly with a regular aquaculture system, it had some disadvantages. It needs water exchange which is a lengthy and costly process.

M. I. Dzulqornain et al. implemented an innovative IoT-based system on the IFTTT model [7]. They used dissolved oxygen, water temperature, the potential of hydrogen (pH) as parameters. The water level was sensed with sensors, and for controlling the system, an aerator system was utilized integrating with microcontroller NodeMCU v3, relay, power supply, and propeller. The sensor data was uploaded to the cloud, and the client could visualize them from anywhere. The application was both web-based and android-based. The system was well enough, but its process was manual.

A. A. Nayan et al. worked on measuring river water quality for agriculture and fishing purposes [8] and identified fish diseases by detecting the changes in water quality [9]. They used a machine learning technique that evaluated the water quality and processed intelligent suggestions. They utilized pH, DO, BOD, COD, TSS, TDS, EC, PO43-, NO3-N, and NH3-N to calculate the water quality and predicted the result using boosting technique. But the study only focused on big water sources like rivers and did not suggest any solution for small water sources like ponds.

To minimize the gap of previous studies and to provide a better understanding of the current state of the art of DL in aquaculture, we have worked on this project. The work offers good support for deploying applications for smart fish farming which is entirely new compared with other results. An automatic system with Biofloc technology has been introduced. We have tried to decrease feed costs by reducing FCR (feed conversion ratio). The nutrients used in this technology can be recycled and reused easily.

## III. MATERIALS AND METHODS

### A. Biofloc Technology (BFT)

Biofloc is a new technology introduced in aquaculture for low-cost fish farming. Bioflocs are used to make reusable food from organic waste nutrients. It is a thin layer made up of beneficial bacteria, microorganisms, and algae that filters the water. Bacteria is cultured for this technology because it produces flocs or algae, breaking ammonia to minimize water pollution. The biofloc method of fish farming can be helpful to grow vegetables and fish together in the yard. This method requires tanks, oxygen supply pumps, and round-the-clock electricity. It needs less amount of food and reduces the chances of the disease. Being an eco-friendly system, it reduces the impact on the environment and improves productivity. Water must be exchanged to minimal or zero in this system. It is cost-effective by reducing the usage of protein-rich feed [10, 11, 12]. Fig. 1 shows a general image captured from a biofloc farm.

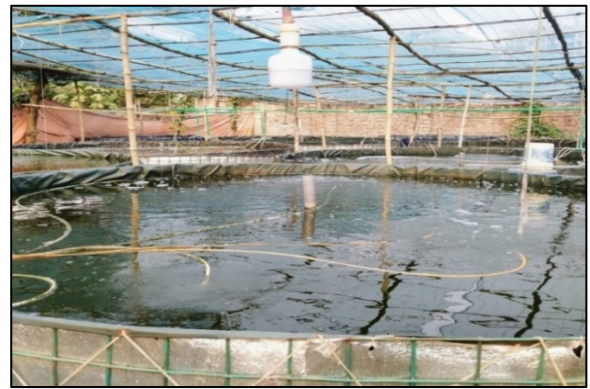

Fig. 1. Biofloc Technology in Aquaculture.

### B. Hardware Components

IoT innovation has brought "Sensor Development" to a new stage. IoT systems operate and use a range of sensors to provide different kinds of information and data. It helps to gather information, drive, and distribute it to a network of similar gadgets. The collected data allows it possible for the devices to work autonomously, and every day the whole world turns to be "smarter." By integrating sensors, microcontrollers, and other smart gadgets, the project was implemented. We have used the following hardware components to run the project.

- Arduino UNO [13].
- White Breadboard.
- pH sensor [14].
- Temperature Sensor [15].
- Total Dissolved Solids (TDS) Sensor [16].
- Computer.
- Wires.

### C. System Architecture

Fig. 2 shows the architecture of our Smart Aquaculture Water Monitoring System. The system monitors continuously and sends notifications through a Wi-Fi [17] module to an android application. The project's primary function is to check the water parameters: pH, temperature, and TDS. We have collected samples from different farms that use Biofloc technology. Processing data from the samples, we trained the artificial neural network. The sensors connected with the system provide a continuous reading of the water parameters. The system generates output evaluating the trained data and the reading provided by the sensors. It predicts the water quality, determines the situation, and process wise decision. It sends the result and decision as a notification to the user through an android application. Fig. 3 shows the flow diagram of the project.

### D. Hardware Connection

We collected the required hardware, tools and connected them according to the diagram shown in Fig. 4. The figure mentions the pin diagram of Arduino Uno with Temperature sensor and pH sensor.

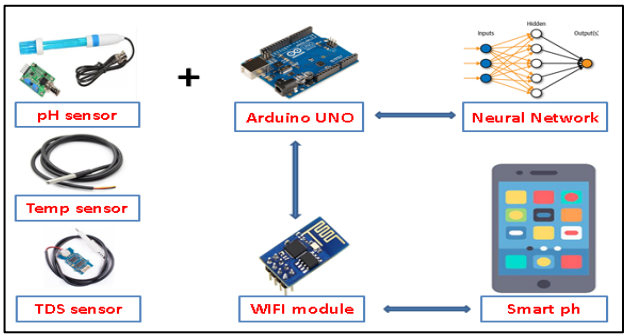

Fig. 2. System Architecture.

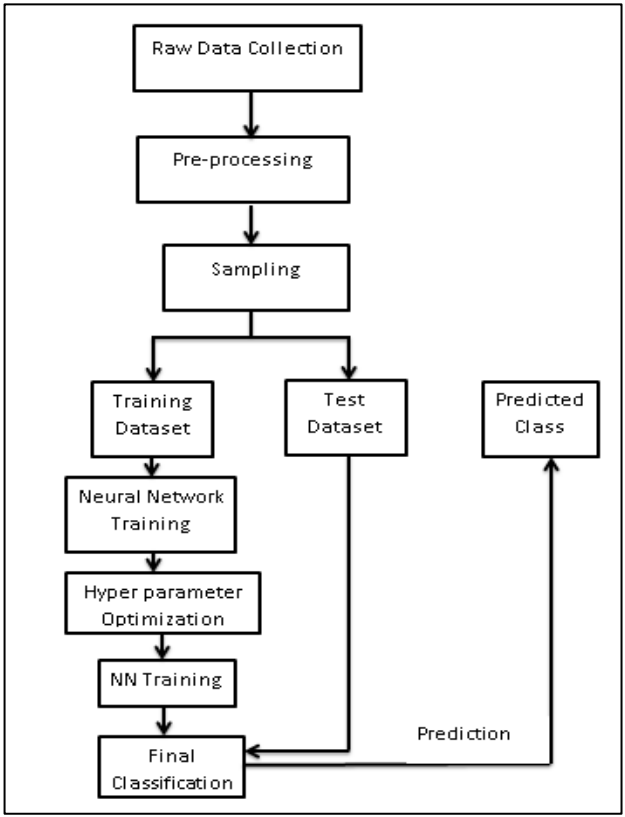

Fig. 3. Flow Diagram.

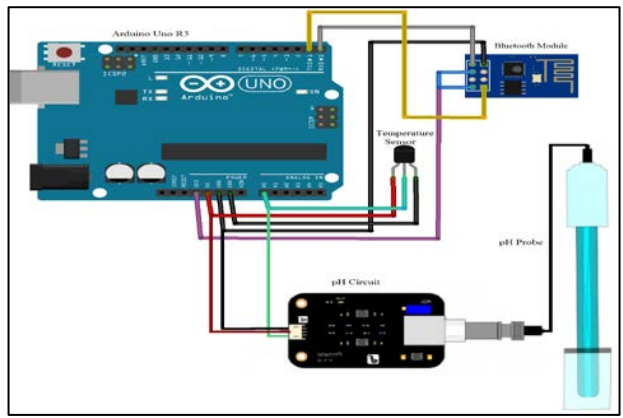

Fig. 4. Hardware Connection.

### E. Study Area

Few farms have already adopted biofloc technology. We found many projects in Bangladesh and visited there. The first center named "Biotech Aquaculture" introduced the technology in Bangladesh. The center is situated at Dakshinkhan, Uttara, Dhaka where Tilapia, Golsha, Pabda, Koi, and Shing (Bengali name of the fishes) are cultivated. The project has been established on the "No Water Exchange" principle. We visited there in September 2020. Investigating their working procedure, we found that they examine the water two times every day and collect the pH, water temperature, total dissolved oxygen (TDS), ammonia (NH3), and floc (molasses). Fig. 5 shows a complete picture of the study area.

Suppose the water parameters get out of the suitable range. In that case, they control them by taking necessary steps for example: filtering out the excess biofloc (as it tends to increase generally), adding baking soda in a safe amount, and removing the fish from the tank before raising the pH. The tank's water is matured, so the parameters do not change frequently, and the water is adaptable for the fish species.

Another center is located at Bosila, Dhaka named "Matsabid Biofloc Aquaculture Farm". They are using the same technology but do not follow the "No Water Exchange" method. It is a big project of 15 large tanks and two large ponds. The whole system is continuously monitored manually by a family residing there. The aerator is mandatory for the biofloc to survive, so the aerator is turned on 24 hours a day. Only three water parameters (pH, salinity, and the number of flocs) are monitored in the farm and whenever one gets increased or decreased, they change the water. For this exchange, they have a drainage system and a water pump. So, the water is not matured here. The image of the study area is shown in Fig. 6 and 7.

### F. Data Collection

We collected samples from different centers. We bought instruments and sensors for data collection purposes. We used pH, temperature, and TDS sensors to measure the values. Sample collection was more manageable, but the data processing was difficult and time-consuming. We worked for more than three months to process the necessary data from the samples. Tables I and II show the details of the collected data.

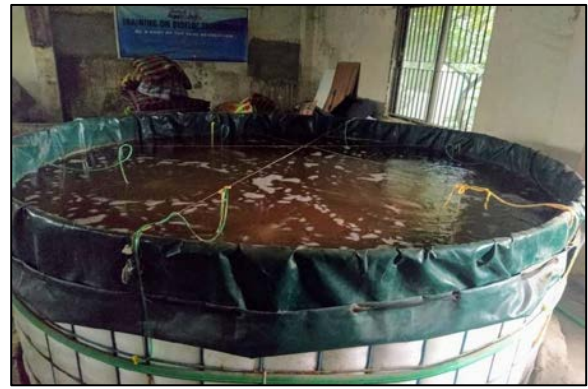

Fig. 5. Study Area 1 at "Biotech Aquaculture".

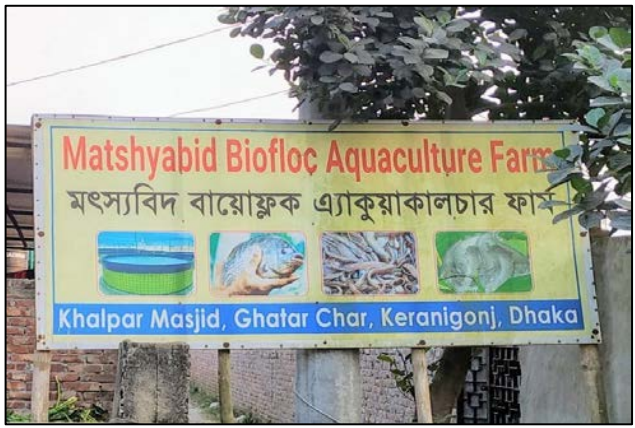

Fig. 6.   Study Area 2 at "Matshyabid Biofloc Aquaculture Farm"

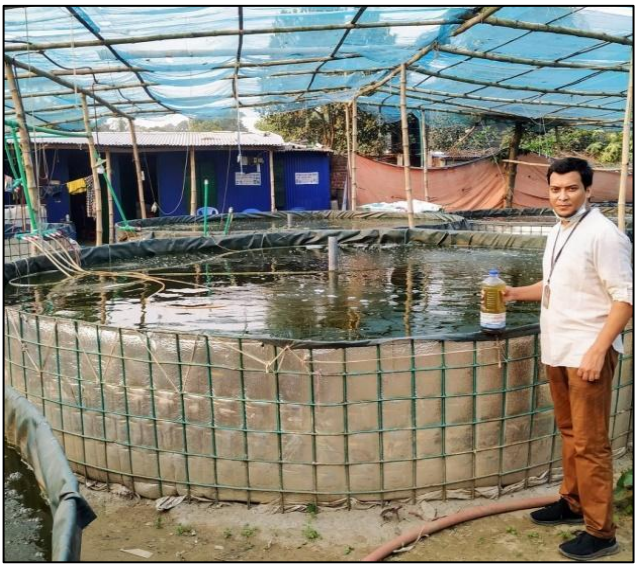

Fig. 7.   Fish Tank at "Matshyabid Biofloc Aquaculture Farm".

TABLE I.    COLLECTED DATA FROM STUDY AREA 1

| Day | Date | pH | Temp | TDS | NH3 | Floc |
|---|---|---|---|---|---|---|
| 1 | 08/06/20 | 7 | 30 | 1.75 | 2 | 10 ml |
| 2 | 09/06/20 | 7 | 30 | 1.6 | 4 | 20 ml |
| 3 | 10/06/20 | 7 | 28 | 1.3 | 2 | 8 ml |
| 4 | 11/06/20 | 7 | 29 | 1.35 | 2 | 10 ml |
| 5 | 12/06/20 | 7 | 28 | 1.32 | 0.25 | 13 ml |
| 6 | 15/06/20 | 7 | 28 | 1.32 | 1 | 10 ml |
| 7 | 16/06/20 | 7 | 30 | 1.59 | 0.25 | 8 ml |
| 8 | 18/06/20 | 7 | 29 | 1.58 | 2 | 10 ml |
| 9 | 19/06/20 | 7 | 28 | 1.33 | 2 | 20 ml |
| 10 | 20/06/20 | 7 | 28 | 1.56 | 0.5 | 15 ml |
| 11 | 21/06/20 | 7 | 30 | 1.60 | 1 | 20 ml |
| 12 | 22/06/20 | 7 | 30 | 1.55 | 0 | 12 ml |
| 13 | 23/06/20 | 7 | 29 | 1.54 | 0 | 20 ml |
| 14 | 25/06/20 | 7 | 30 | 1.75 | 0 | 8 ml |

TABLE II.    COLLECTED DATA FROM STUDY AREA 2

| Date | Time | pH | TDS | Floc |
|---|---|---|---|---|
| 18/10/20 | 7.00 am | 8.2 | 652 | 45 gm |
| 18/10/20 | 8.00 pm | 8.2 | 684 | 55 gm |
| 19/10/20 | 7.00 am | 8.2 | 684 | 43 gm |
| 20/10/20 | | 8.3 | 684 | 54 gm |
| 21/10/20 | 7.00 am | 8.3 | 659 | 22 gm |
| 21/10/20 | 8.00 pm | 8.4 | 654 | 25 gm |
| 22/10/20 | 7.00 am | 8.5 | 682 | 25 gm |
| 22/10/20 | 8.00 pm | 7.1 | 678 | 50 gm |
| 23/10/20 | 7.00 am | 7.9 | 654 | 45 gm |
| 25/10/20 | 9.00 am | 8.3 | 658 | 30 gm |
| 26/10/20 | 7.00 am | 8.3 | 684 | 29 gm |
| 26/10/20 | 8.00 pm | 8.4 | 681 | 35 gm |
| 27/10/20 | 9.00 am | 7.9 | 652 | 30 gm |
| 28/10/20 | 7.00 am | 8.3 | 685 | 33 gm |
| 28/10/20 | 8.00 pm | 8.2 | 682 | 45 gm |

*G. Data Preprocessing*

We collected data from water using different instruments and stored it in a CSV file. The CSV file contains six various labels, including temperature, TDS, pH, and flocs. pH is considered as the output data by which the model was tested and the rest five are regarded as the input data by which the model was trained. A glimpse of our dataset is shown in Table III.

*H. Machine Learning Algorithm for Decision Making*

Artificial Neural Networks (ANN) function as the neurons of the human brain. The network contains nodes that receive the input signal and pass it to the previous nodes as the synapse of a nerve cell does [18, 19]. The covariates and input variables are weighted, and these weighted signals are then passed through activation functions. Let y be output signal and be the activation function, the mathematical expression of signal processing in ANN is:

$$y(x) = \Phi(\sum i = 1 \; w\, i \cdot x\, i)$$

The network contains an input layer, an output layer, and one or more hidden layers. The hidden layers are responsible for the performance of the model. We used five hidden layers for faster execution. More layers slow down the training and testing process.

*I. Dependencies*

We trained the model on Ubuntu 20.04 LTS and used python 3.8. These were dependencies and libraries. We installed and imported the following libraries to run the code and train the model.

- Tensor flow

- Keras

- Pandas

- Matplot
- Numpy
- Boxplot Analysis

### J. Parameters and Algorithms

The whole dataset was splatted into two different parts that are training and testing. We took 80% of the data for training, and the rest 20% was used for testing purposes. The model was trained several times using different epoch sizes. We encountered the overfeeding condition. Lastly, utilizing batch size 32 and epoch size 150, the model achieved the best accuracy. The model consists of 5 hidden layers. The layers help to increase the model's performance. The workflow of the layers is shown in Fig. 8.

To vanish the gradient problem and allow the model to run faster and perform well, we used ReLu (Rectified Linear Unit) [20, 21] with the hidden layers. Softmax was utilized as an activation function in the output layer. The number of classes was 4 for the output [22]. 0 denotes a shallow DO level, 1 represents a low DO level, 2 denotes an average DO level, and 3 denotes a high DO level.

TABLE III. PROCESSED DATA

|   | A | B | C | D | E |
|---|---|---|---|---|---|
| 1 | Temp | D.O. (mg) | pH | TDS | Flocs (ml) |
| 2 | 29.5 | 6.3 | 6.9 | 1.7 | 10 |
| 3 | 29.7 | 5.7 | 6.9 | 3.8 | 50 |
| 4 | 29.5 | 5.8 | 7.3 | 1.9 | 40 |
| 5 | 30 | 5.5 | 7.4 | 1.5 | 10 |
| 6 | 29.2 | 6.1 | 6.7 | 1.4 | 30 |
| 7 | 29.1 | 7.3 | 7 | 1 | 10 |
| 8 | 28.7 | 7 | 6.9 | 1.2 | 30 |
| 9 | 28.7 | 7.3 | 6.7 | 1.5 | 10 |
| 10 | 29.5 | 7.2 | 6.8 | 1.2 | 30 |
| 11 | 29 | 5.3 | 6.4 | 1.6 | 120 |
| 12 | 30.5 | 6.3 | 7.5 | 1.5 | 10 |
| 13 | 29.1 | 5.5 | 6.3 | 1.4 | 10 |
| 14 | 30.1 | 7.3 | 7.8 | 2 | 30 |
| 15 | 29.2 | 6.5 | 7.9 | 1.5 | 40 |
| 16 | 25.1 | 7.2 | 7.7 | 4.9 | 30 |
| 17 | 29.6 | 6.6 | 7.8 | 5 | 60 |
| 18 | 27.4 | 6.9 | 7.3 | 5.2 | 30 |
| 19 | 27.8 | 6.8 | 7.9 | 4.9 | 180 |
| 20 | 30.6 | 6.7 | 7.6 | 10.3 | 190 |
| 21 | 25 | 5.1 | 7.6 | 3.6 | 60 |
| 22 | 28.1 | 5.6 | 7.7 | 4.6 | 70 |
| 23 | 28.6 | 6.3 | 6.9 | 4.7 | 40 |
| 24 | 26.9 | 6.6 | 6.8 | 5 | 30 |
| 25 | 28.2 | 6.8 | 6.5 | 5.2 | 60 |

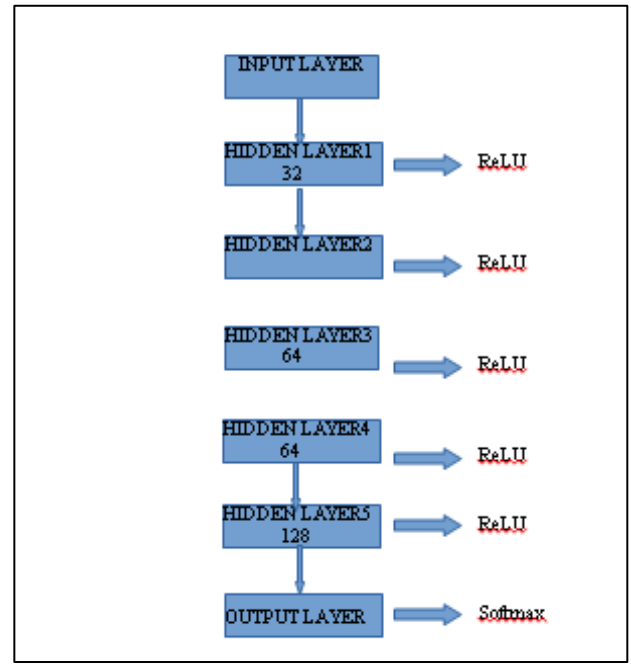

Fig. 8. Workflow of the Layers.

## IV. SIMULATION AND RESULT

### A. Training and Testing

After collecting and processing the dataset, we trained the model. The model was trained with the 80% data and later tested with the rest 20% data. We utilized a completely different type of data for training and testing purposes. The model scored 0.773 testing accuracy, which was well enough to maintain good performance. The loss was calculated with the increasing number of epochs. After 55 epochs, the loss was minimized rapidly. We calculated the loss between 0 to 1.2 range. The minimum testing loss was 0.5, and the training loss was 0.7. The training and testing accuracy and loss have been shown in Fig. 9 and 10.

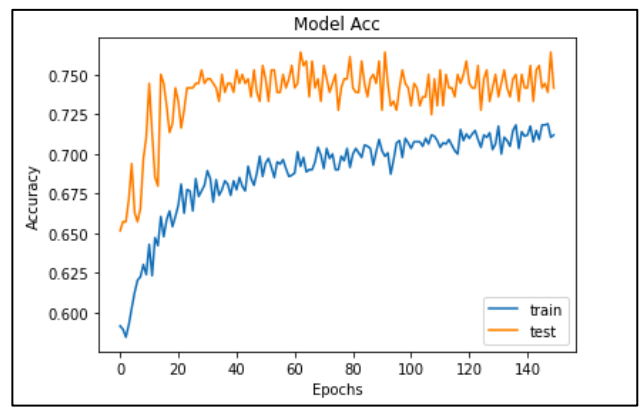

Fig. 9. Training and Testing Accuracy.

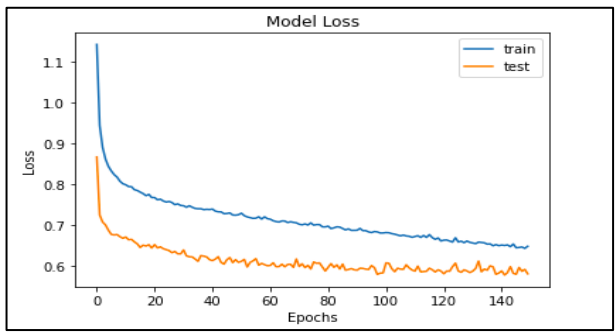

Fig. 10. Training and Testing Loss.

### B. System Output

Evaluating the parameter's value, the system provides an output. Depending on the DO level (shallow, low, average, or high) in the water, the system generates a message and provides a decision analyzing the current output. To make the system more user-friendly, we have designed a smartphone application to notify the user about the result and determination that the machine generates.

A user needs the android application, an ubuntu droplet, and a droplet's IP address for the push notification. We used GCM (Google Cloud Messaging) [23] for android, where we enabled the API for our project first and then linked the android App through it. We deployed an Ubuntu droplet and set up a python GCM simple server on Ubuntu. Lastly, a push notification was displayed on the Android app generated by the system. A snapshot of the smartphone application is shown in Fig. 11.

### C. Performance Comparison

Many researchers have worked on intelligent biofloc technology. pH, DO, BOD, COD, TSS, TDS, EC, PO43-, NO3-N, and NH3-N are the standard parameters utilized by most researchers for measuring water quality and its changes. The Artificial Neural Network (ANN) [24], Group Data Handling Method (GMDH) [25], Support Vector Machine (SVM) [26], Least-Squares Support Vector Regression (LSSVR) [27] and Long Short-Term Memory (LSTM) [28] are commonly used methods. Besides, some projects used microcontrollers and sensors only. The projects are cost-effective, but such projects do not predict the water quality, fish health condition and cannot process automatic solutions. A caretaker is needed to perform the manual functions. In this article, we have proposed and implemented a system that will help fish farmers grow fish production minimizing the feeding cost. It needs not to use big ponds or a wider area. Our project can help anyone to produce plenty of fishes in a small container or house at a small cost. We have compared our approach with other existing techniques. For comparing, we provided significant importance to the proposed method, accuracy, cost reduction rate, parameters, real-time monitoring capability, prediction capability, decision-making capability, and user satisfaction level. The comparison is shown in table 4. The information mentioned in the table is collected from different published articles. Here we did not compare among the methods. We have reached the performances of different approaches only.

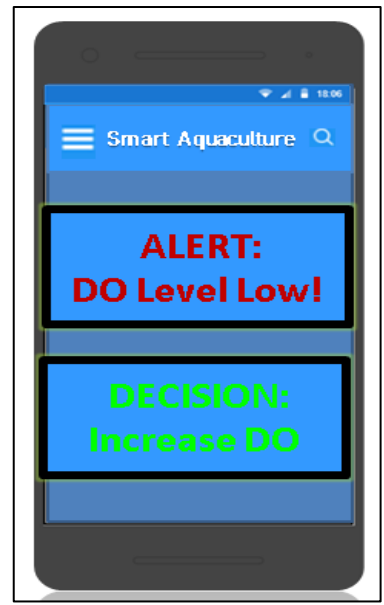

Fig. 11. Notification to the Smartphone Application.

TABLE IV.    PERFORMANCE COMPARISON

| Method Used | Testing Accuracy (Percentage) | Real-time Monitoring | Parameters | Automatic Solution | Decision-Making Ability | User satisfaction |
|---|---|---|---|---|---|---|
| AAN | 72 % | Monitors 24 x 7 | Temperature, DO, TDS, pH, BOD, COD, TSS | Does not perform automatic solution | Can predict and process smart decision | Medium |
| GMDH | 74% | Monitors 24 x 7 | Temperature, DO, TDS, pH | Does not perform automatic solution | Better prediction and decision-making system | High |
| SVM | 70% | Monitors 24 x 7 | Temperature, DO, TDS, pH | Does not perform automatic solution | Lower quality prediction | Low |
| LSSVR | 76% | Monitors 24 x 7 | Temperature, DO, TDS, pH, EC, PO43-, NO3-N | Does not perform automatic solution | Can predict and process smart decision | Medium |
| LSTM | 82 % | Monitors 24 x 7 | pH, temperature, DO | Does not perform automatic solution | Better prediction and decision-making system | High |
| **Our Approach** | 77 % | Monitors 24 x 7 | Temperature, DO, TDS, pH, Floc | Performs automatic solution | Can predict and process smart decision | Android application is available for monitoring from anywhere. User satisfaction is measured as "Very High" |

## V. Conclusion

Aquaculture farmers have been surviving from economic constraints, high-paid and even unavailability of human resources, timely monitoring of water quality, and sudden increase in toxicity for decade after decade. The IoT-based water quality monitoring system monitors the water quality in real-time and reduces the cost of production, increases efficiency, reduces human dependency, and thus ensures sustainable development economically and socially. The proposed system monitors the water quality in real-time and sends a notification to the user instantly, which reduces the risk. A machine learning technique has been applied to trace the water quality. To validate the proposed model, experiments on the implemented functionalities have been performed. The experiments show 0.773 as testing accuracy which was well enough to maintain good performance. Currently, the implementation of identified functionalities has been carried out. In the future, we wish to improve the model to achieve higher accuracy and evaluate the performance in terms of the fish population.

## Acknowledgment

The IoT Lab has supported this research work, a state-of-art specialized research facility of its kind situated at ULAB, implemented, and supported by the Bangladesh Hi-Tech Park Authority of ICT Division.